\documentstyle[epsfig]{mn2e}

\title{Voids in a $\Lambda$CDM Universe}
\author[J.\ M.\ Colberg {\it et al.}]
       {J\"org M.\ Colberg,$^1$ Ravi K.\ Sheth,$^1$ Antonaldo Diaferio,$^2$ 
        Liang Gao,$^3$ and \newauthor Naoki Yoshida$^4$\\
        $^1$ University of Pittsburgh, 3941 O'Hara Street, 
             100 Allen Hall, Pittsburgh PA 15260, USA\\
        $^2$ Università degli Studi di Torino, Dipartimento di Fisica Generale 
             Amedeo Avogadro, Via Pietro Giuria 1, 10125 Torino, Italy\\
        $^3$ Max--Planck-Institut f\"ur Astrophysik, 
             Karl--Schwarzschild--Str. 1, 85741 Garching, Germany\\
        $^4$ Department of Physics, Nagoya University, Nagoya, 
             Aichi 464--8602, Japan}
\date{Accepted 200? ???? ??.
      Received 2004 ???? ??;
      in original form 2004  xx}

\pagerange{\pageref{firstpage}--\pageref{lastpage}}
\pubyear{200?}

\begin{document}

\maketitle

\label{firstpage}

\begin{abstract}
We study the formation and evolution of voids in the dark matter distribution using various 
simulations of the popular $\Lambda$ Cold Dark Matter cosmogony. We identify voids by 
requiring them to be regions of space with a mean overdensity of $-0.8$ or less -- roughly 
the equivalent of using a Spherical Overdensity group finder for haloes. Each of the 
simulations contains thousands of voids. The distribution of void sizes in the 
different simulations shows good agreement when differences in particle and grid 
resolution are accounted for. Voids very clearly correspond to minima in the smoothed initial 
density field. Apart from a very weak dependence on the mass resolution, the rescaled mass 
profiles of voids in the different simulations agree remarkably well. We find a universal 
void mass profile of the form  $\rho(<r)/\rho(r_{\rm eff}) \propto \exp[(r/r_{\rm eff})^\alpha]$ 
where $r_{\rm eff}$ is the effective radius of a void and $\alpha\sim 2$. The mass function 
of haloes in voids is steeper than that of haloes that populate denser regions. In addition, 
the abundances of void haloes seem to evolve somewhat more strongly between redshifts $\sim 1$ 
and 0 than the global abundances of haloes.
\end{abstract}


\section{Introduction}

Galaxy redshift surveys show that galaxies are not distributed uniformly. Instead, they form 
a complicated network around large regions that are almost empty, so-called voids. One of the 
most famous voids, in the region of Bo\"otes, has a diameter of $\sim 50h^{-1}$Mpc, and was 
found by Kirschner et al (1981).\footnote{Throughout this work, we will express the Hubble 
constant in units of $H_0 = 100 h\,$km/sec/Mpc.} Subsequent larger redshift surveys found more 
and more voids (for example Geller \& Huchra 1989, da Costa et al 1994, Shechtman et al 1996, 
Einasto et al 1997, Plionis \& Basilikos 2002). These surveys allowed studies of the properties 
of voids and of void galaxies (Einasto et al 1994, Lindner et al 1995 and 1996, El--Ad et al 
1997, M\"uller et al 2000), but only recently have galaxy surveys become large enough to yield 
sufficient sample sizes for systematic studies (Hoyle \& Vogeley 2002, 2004; Rojas et al 2003; 
Croton et al 2004).

For similar reasons, voids in cosmological N--body simulations have also been less well--studied. 
Early simulations of Cold Dark Matter (CDM) universes showed that large empty regions were generic 
(Icke 1984, Davis et al 1985), and larger more recent simulations (e.g., Jenkins et al 1998) have 
provided a clearer picture of the `void hierarchy' (Van de Weygaert \& Van Kampen 1993, Sheth \& 
Van de Weygaert 2004). Detailed studies of the properties of voids in the dark matter distribution 
are now becoming increasingly common (Little \& Weinberg 1994, Gardner 2001, Schmidt et al 2001, 
Gottl\"ober et al 2003, Patiri et al 2004). 

Peebles (2001) noted that the properties of CDM voids and of the galaxies inside them formed a 
strong test for CDM. Subsequently, Mathis \& White (2002) and Benson et al (2003) investigated 
properties of voids in semi--analytical models where mock galaxies are placed in dark--matter 
only simulations following physically motivated recipes.

One of the problems with voids and with studies of voids is that there is little agreement on 
how to define a void in the galaxy distribution. Are voids regions which are completely devoid of 
galaxies? Or can there be galaxies inside a void? If yes, how do void galaxies differ from their 
cousins that populate denser environments? And what is the spatial distribution of void galaxies 
within voids? Are they scattered throughout the void interior, or do they tend to pile up around 
the edges? 

In models of galaxy formation within the context of hierarchical clustering, the galaxy distribution 
is determined by the underlying dark matter.  Therefore, to understand void galaxies, it is important 
to define precisely what constitutes a void in the dark matter distribution. Dubinski et al (1992) 
argued that the spherical evolution model (Gunn \& Gott 1972) provides a useful guide. In this model, 
initially underdense regions evolve from the inside out, in the sense that as mass makes its way 
outwards from the centre of the underdensity, a well--defined ridge begins to form around the region.  
The ridge is well--formed at a time when the density interior to it has a characteristic value (e.g. 
Fillmore \& Goldreich 1984; Bertschinger 1985), and this, they argued, provides a natural and physically 
motivated definition of a void (also see Van de Weygaert \& Van Kampen 1993; Friedmann \& Piran 2001; 
Sheth \& Van de Weygaert 2004). The main purpose of the present work is to present the results of a 
study which uses this definition of voids.  

We will concentrate on voids in what has now become the standard cosmological model: a flat $\Lambda$CDM 
cosmology with $\Omega=0.3$. Rather than focusing on individual voids, or small sets of voids, we take 
a series of high--resolution N--body simulations done in sufficiently large cosmological volumes 
that a study the properties of ensembles of voids is justified. The set of simulations we use covers 
a wide range of cosmological volumes and resolutions. Thus, we are able to study detailed properties 
of voids such as the density run of the matter within them, as well as estimate their abundances. 

This paper is organized as follows. In the following Section, we introduce the simulation set 
(\S~\ref{simulations}) and the void finding algorithm (\S~\ref{voidfinder}). In Section~\ref{voids} 
we study the visual appearance of voids (\S~\ref{visual}), the void volume function 
(\S~\ref{VoidVolumeFunction}), the correspondence between voids and minima in the initial density 
field (\S~\ref{troughs}), density profiles of voids (\S~\ref{profiles}), the mass function of haloes 
in voids (\S~\ref{massfunction}), and the spatial clustering of voids (\S~\ref{corrfct}). 
Section~\ref{conclusions} summarizes our findings.

\section{Finding Voids in Cold Dark Matter Universes}

\subsection{The Simulations} \label{simulations}

We use of a set of N--body simulations done by, or in collaboration with, the Virgo Supercomputing 
Consortium\footnote{http://www.virgo.dur.ac.uk}. The simulations model regions of different sizes and 
have different mass resolutions.  In the naming conventions of the Virgo Consortium, the simulations 
are: 
\begin{enumerate}
  \item The $\Lambda$CDM GIF simulation (Jenkins et al 1998; Kauffmann et al 1999), with $256^3$ 
        particles in a cubic volume of size $(141.3\,h^{-1}\,$Mpc)$^3$. 
  \item The $\Lambda$CDM GIF2 simulation (Gao et al 2003), with $400^3$ particles in a cubic volume of 
        size $(110\,h^{-1}\,$Mpc)$^3$.  The mass resolution of this simulation is ten times better 
        than that of the GIF simulation. 
  \item The $\Lambda$CDM VLS simulation (Jenkins et al 2001; Yoshida et al 2001; Menard et al 2003), 
        with $512^3$ particles in a cube of volume $(479\,h^{-1}$\,Mpc)$^3$. This simulation has the
        same mass resolution as the largest boxes in Jenkins et al (1998) but is eight times their
        volume.
  \item The $\Lambda$CDM Hubble Volume simulation (Evrard et al 2002), with $1000^3$ particles in a  
        $(3000\,h^{-1}\,$Mpc)$^3$ cube. Despite the relatively low mass resolution of this simulation, 
        its size makes it extremely useful for studying the largest possible voids.  
\end{enumerate}
We note a difference in the initial power spectra for these simulations. The initial condition for
the GIF simulation was generated using the Bond \& Efstathiou (1984) transfer function
whereas for the other simulations the transfer function computed by CMBFAST (Seljak \& Zaldarriaga 1996)
for the LCDM model was used. Table~\ref{simtable} provides a few more details about the simulations; in 
the following, we will refer to them as GIF, GIF2, VLS, and HV.

\begin{table}
  \begin{center}
  \begin{tabular}{lrrr}
    \hline
     Run & $n_p$ & $l$ [$h^{-1}$\,Mpc] & $m_p [10^{10}\,h^{-1}\,M_{\odot}]$\\
    \hline
     GIF & $256^3$ & 141.3 & 1.4\\
     GIF2 & $400^3$ & 110 & 0.2\\
     VLS & $512^3$ & 479 & 6.9\\
     HV & $1000^3$ & 3000 & 224.8\\
    \hline 
  \end{tabular}
  \end{center}
  \caption{Parameters of the simulations used in this work. All runs have $\Omega_m=0.3$, $\Lambda=0.7$, 
           $n=1$, $\sigma_8=0.9$, $h=0.7$.}
  \label{simtable}
\end{table}

\begin{figure}
\caption{File {\bf http://lahmu.phyast.pitt.edu/$\sim$colberg/voids/criteria.gif} -- 
         Merging criteria for proto--voids: (a) Void 2 lies fully inside void 1 and thus belongs to void 1.
         (b) Void 2's center lies inside void 1; the resulting void consists of void 1 plus the additional 
         volume of void 2 that lies outside of void 1. (c) Void 2's center lies outside of void 1, but the
         region of overlap is large enough to make the algorithm merge the two voids: 'x' is longer than 
         both 'y' and 'z'. (d) If void 2 was merged with void 1 then the algorithm will not look whether
         it also overlaps with void 3.}
\label{criteria}
\end{figure}

\begin{figure}
\includegraphics[width=85mm]{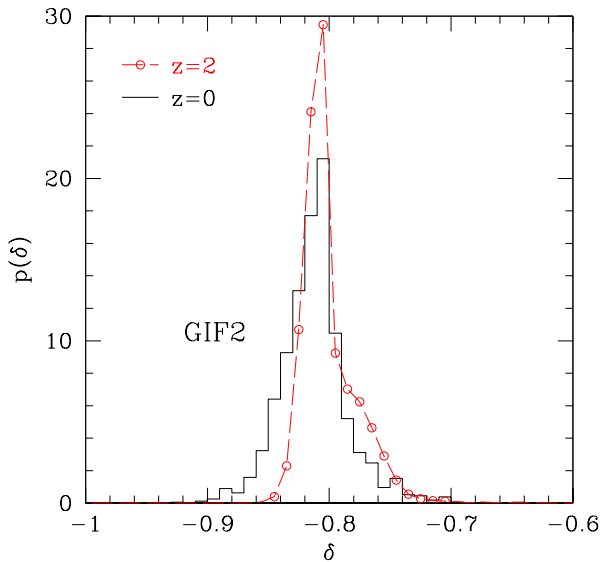}
\caption{Distribution of (merged) void overdensities in the GIF2 simulation at redshifts $z=0$ (solid 
         histogram) and 2 (connected circles). The overdensities scatter around the value of -0.8 used
         for the selection of the proto--voids (see description of void finding algorithm for explanation
         and discussion).}
\label{GIF2deltas}
\end{figure}

\begin{figure}
\includegraphics[width=85mm]{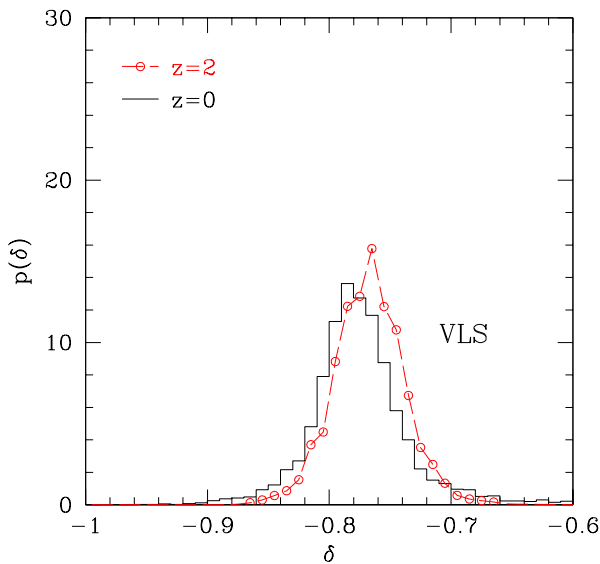}
\caption{Distribution of (merged) void overdensities in the VLS simulation at redshifts $z=0$ (solid 
         histogram) and 2 (connected circles). The overdensities scatter around a value of slightly 
         less than -0.8 used for the selection of the proto--voids (see description of void finding 
         algorithm for explanation and discussion).}
\label{VLSdeltas}
\end{figure}

\subsection{The Void Finding Algorithm} \label{voidfinder}

A number of void-finding algorithms have been proposed (Kauffmann \& Fairall 1991, Kauffmann \& Melott 
1992, El-Ad \& Piran 1997, Aikio \& M\"ah\"onen 1998, Hoyle \& Vogeley 2002). Most look for empty spherical 
or cubical regions, which are then merged following some recipe.  For the galaxy distribution, the decision 
to merge or not is slightly ad hoc.  Our task is somewhat easier, because we are only searching for voids 
in the dark matter distribution, and we have a dynamically based model to use as a guide.  

Our void finding algorithm is a variant of the one advocated by Aikio \& M\"ah\"onen (1998). It is based 
on the assumption that voids are primordial negative overdensity perturbations that grew gravitationally 
and have reached shell--crossing at present time. At shell-crossing, the comoving radius of a perturbation 
is $1.7$ times larger than it was initially, so that the object has a density contrast of $-0.8$ (see 
Blumenthal {\it et al.} 1992, Dubinski {\it et al.} 1993). (Strictly speaking, these numbers are correct 
for an Einstein de-Sitter cosmology.  But the dependence on cosmology is weak, and so we ignore it.)  
Our algorithm looks for such regions in the simulations. To be more precise, it performs the following steps:
\begin{enumerate}
  \item The simulation particles are binned on a three--dimensional mesh using a nearest gridpoint scheme. 
        We have checked that the choice of the grid size does not influence the locations and sizes of the
        voids, provided the smallest voids have radii of at least three cells.
  \item The grid is smoothed adaptively, using 20 particles for the smoothing kernel. A fixed smoothing 
        filter, for instance a Gaussian of some radius, smoothes the relatively large underdense regions 
        nicely but washes out the smaller, highly clustered regions. As noted in the previous section, 
        previous studies indicate that voids have well-defined steep edges which a fixed smoothing would 
        wash out. The adaptive smoothing ensures that only regions with small particle numbers are heavily 
        smoothed.  
  \item Local minima in the particle distribution are found, and spheres of varying radii are centred on 
        these minima. The overdensity within these spheres is computed. The largest sphere within which 
        the overdensity is $-0.8$ (or slightly smaller) is added to our list of void building blocks.  
  \item In principle, the entire underdense volume in a simulation box forms one big interconnected void: 
        collapsed haloes form very small islands of matter surrounded by a vast underdense ocean
        (recall that collapsed haloes are much denser than the background, so they occupy only a small 
        fraction of the volume). We divide this underdense ocean up into smaller voids that we require 
        to either be spherical or ellipsoidal, or to have any other irregular shape, provided that they 
        do not consist of two (or more) regions connected by thin tunnels. To avoid dumbbell--shaped 
        configurations, the spherical void building blocks are merged using the following criteria (c.f.\
        Figure~\ref{criteria}):
        \begin{enumerate}
          \item Any sphere which lies fully inside another is eliminated from the list.
          \item Any smaller sphere whose centre lies within a larger sphere is considered to be part of the 
                larger volume; the void then contains the volume of the first sphere plus the additional 
                volume of the second sphere. The overdensity of the resulting void is computed using its
                volume and the matter it contains.
          \item A sphere whose center lies outside the boundary of another sphere is considered to be part 
                of the other if the following requirements are met. First, the two spheres must overlap. 
                Second, the line which connects the centers of the two spheres is divided into three 
                segments. A central part, which lies within the volume of intersection of the two spheres, 
                and the two ends which do not. If the central segment is longer than one of the other two, 
                the two spheres are considered to be part of the same void.  
          \item If a sphere overlaps with another sphere the merging algorithm will not look for more 
                overlaps. Thus, two large voids will never be connected by a thin bridge because the 
                algorithm places a small sphere in between them only into one and not into both voids. 
                This way, dumbell--shaped configurations are not possible. 
        \end{enumerate}
\end{enumerate}

Our void finder is analogous to the spherical overdensity method for dark matter haloes (Lacey \& Cole 1994). 
We compute the center of each void by taking the volume--weighted average of the centers of its constituent 
spheres. By construction, voids need not be spherical, and we do not attempt to quantify the geometric shapes 
of the voids any further (for a discussion of this subject see Van de Weygaert \& Van Kampen 1993 and
references therein). Instead, we compute an `effective' radius by taking the radius of a sphere whose 
volume is equal to that of the void. The effective radius has no deeper physical meaning but it is quite useful 
to get some idea of how big a void actually is. But note that in the spherical evolution model, the initial 
spherical region from which the void grew differs from this effective radius by a factor of $(1+\delta)^{1/3}$.  

By running our algorithm on the simulated particle distributions at different epochs, we obtain void samples 
at a range of redshifts. We only consider voids whose radii are four times larger than the scale of the 
void--finder grid. Table~\ref{paramstable} summarizes our results. The void radii/sizes in the different 
simulations are discussed in more detail in Section \ref{VoidVolumeFunction}.  

Figures~\ref{GIF2deltas} and \ref{VLSdeltas} show the overdensities of the (merged) voids in the GIF2 and VLS 
simulations, respectively, at redshifts $z=0$ and $z=2$. The distributions scatter nicely around the value 
$\delta=-0.8$, with the peaks of the GIF2 and VLS distributions slightly above or below $\delta=-0.8$, 
respectively. This slight difference is due to the somewhat coarser grid of the VLS simulation.

\begin{table}
  \begin{center}
  \begin{tabular}{lcccr}
    \hline
     Run & $z$ & $r_{min}$ & $r_{max}$ & $n$ \\
         &     & [$h^{-1}$\,Mpc] & [$h^{-1}$\,Mpc] &   \\
    \hline
     GIF  & 0 & 1.2 & 32.1 & 5460 \\
          & 1 & 1.2 & 16.5 & 8597 \\
          & 2 & 1.2 &  9.0 & 5564 \\
          & 3 & 1.2 &  4.3 & 1660 \\
     GIF2 & 0 & 0.7 & 19.8 & 7605 \\
          & 1 & 0.7 & 14.3 & 14331 \\
          & 2 & 0.7 &  6.3 & 21835 \\
          & 3 & 0.7 &  4.3 & 13957 \\
     VLS  & 0 & 3.5 & 33.2 & 46405 \\ 
          & 1 & 3.5 & 15.2 & 45592 \\ 
          & 2 & 3.5 &  9.1 & 11730 \\ 
          & 3 & 3.5 &  5.5 &  1063 \\ 
     HV   & 0 & 10.0 & 55.9 & 77726 \\ 
    \hline 
  \end{tabular}
  \end{center}
  \caption{Void samples from the simulation sets. $r_{min}$ is the lower threshold for the void samples; 
           $r_{max}$ is the effective radius of the largest void in the sample; $n$ denotes the total 
           number of voids larger than $r_{\rm min}$ in our sample.}
  \label{paramstable}
\end{table}

\section{Voids in a $\Lambda$CDM Universe} \label{voids}

\subsection{Visual Impression} \label{visual}

High--resolution N--body simulations contain a large number of three-dimensional objects. The appearance of 
these objects is usually illustrated by plotting the smoothed or unsmoothed particle distribution from a 
narrow slice through the simulation volume.  However, projection effects can make objects seem to lie in 
the wrong places. What is more, images of smoothed density distributions are usually plotted using a 
{\it logarithmic\/} scale which tends to emphasize the matter between the haloes over the haloes themselves. 
(If a linear scale is used, most of the image would be relatively featureless, except for a few tiny specks 
that represent the haloes.)  We will use a logarithmic scale in what follows, but it is important to keep 
this caveat in mind when looking at the images.

\begin{figure*}
\caption{File {\bf http://lahmu.phyast.pitt.edu/$\sim$colberg/voids/GIF2\_LCDM\_voids\_V2.jpg} --
         A slice of thickness $10\,h^{-1}$\,Mpc through the GIF2 simulation. The dark matter was smoothed 
         adaptively, and the resulting density field is shown using a logarithmic colour scale. Circles show 
         the effective radii of each void and are drawn to guide the eye (see main text for a more
         detailed description of how voids whose centers lie inside and outside the slice are represented). 
         Numbers refer to a few points that have to be made about the plot (for more details see the main
         text): (1) Voids are shown as circles but in reality, they are not spherical. These three large 
         voids do not overlap in our void catalog.(2) This part of an underdense region is not covered 
         by any of the voids in this image because they are drawn as circles. In reality, it is part of 
         the large void right above it. (3) Smaller voids seem to lie inside bigger ones. This is due to a 
         combination of projection effects and of the fact that we draw voids as circles. The same goes for 
         large haloes in voids. In the image, we have marked some of those voids that seem to be contained
         inside the larger void. (4) Regions that, on larger scales, are more overdense contain mostly small 
         voids. (5) There are some small haloes inside voids. We have marked some of the small haloes inside
         the large void.}
\label{GIF2slice}
\end{figure*}

Figure~\ref{GIF2slice} shows a slice of thickness $10\,h^{-1}$\,Mpc -- about one tenth of the full box size -- 
through the GIF2 simulation. The dark matter was smoothed adaptively, and the density distribution was plotted 
using a logarithmic colour scale.  The circles superimposed on the density field are centred on the centers 
of those voids that intersect this slice. For voids whose centers lie inside the slice we plot a circle with
a radius equal to the effective radius. For voids whose centers lie outside the slice we determine the size
of the overlap between the slice and the void that we represent as a sphere. We then plot a circle whose
radius corresponds to the radius of the circle that is defined by the intersection of the sphere with the
outer edge of the slice. The figure illustrates that the effective radii defined above correspond quite 
nicely to the visual impression of voids' sizes. We added a few numbers to the image at locations that require 
some attention: (1) For reasons of simplicity, voids are shown as circles. In reality, they are not spherical. 
Thus, these three large voids do not overlap in the void catalog. (2) There is a small region here, which is 
underdense but not covered by any of the voids in this image. This effect is also due to the fact that we draw 
voids using circles. In reality, this underdense region is part of the large void right above it. (3) Smaller 
voids seem to lie inside bigger ones. This does not actually happen in our void catalogue. In the image, it is 
due to a combination of projection effects and of the fact that we draw voids as circles. Larger haloes also 
do not lie inside voids but are merely projected on top of them. It is worth noting that as a result of
projection effects, if the centers of some of the larger voids lie close to the edge of the slice, they appear 
to be much larger than the particle distribution in the slice would have indicated. (4) Note how regions that 
are more overdense do not contain many large voids but, instead, mostly small ones. (5) There are some small 
haloes inside voids as is clearly visible in the center of this very large void.

Figure~\ref{GIFmosaic} shows the growth of the three largest voids in the GIF simulation. We are plotting a slice 
of thickness $10\,h^{-1}$\,Mpc. However, in this plot, we center the region we are plotting on the $z=0$ position 
of each void. The evolution of the voids is quite interesting and it seems to follow the general picture outlined 
above: Even though the actual void shapes are not spherical, the voids grow from the inside, expanding outwards.
Also note the presence of a very large void in the left--most column. This void has almost the size of the
largest void in the VLS simulation.

\begin{figure*}
\caption{File {\bf http://lahmu.phyast.pitt.edu/$\sim$colberg/voids/GIF\_LCDM\_void\_mosaic\_bw\_V2.jpg} --
         The largest voids in the GIF $\Lambda$CDM simulation. Each set of slices shows a 
         $40\times40\times10$\,($h^{-1}$\,Mpc)$^3$ volume centred on one of the three largest voids. The sequence 
         from top to bottom shows each void at $z=0$, 1, 2, and 3. The colour coding is the same for all voids and 
         redshifts. At $z=0$, the three voids have effective radii of 32.1\,$h^{-1}$\,Mpc (leftmost column),
         18.9\,$h^{-1}$\,Mpc (middle column), and 18.7\,$h^{-1}$\,Mpc (rightmost column).}
\label{GIFmosaic}
\end{figure*}

\subsection{The Void Volume Function} \label{VoidVolumeFunction}

\begin{table}
  \begin{center}
  \begin{tabular}{lr}
    \hline
     $z$ & $f$ \\
    \hline
     0 & 61.2\%\\
     1 & 27.6\%\\
     2 & 9.2\%\\
     3 & 2.7\%\\
    \hline 
  \end{tabular}
  \end{center}
  \caption{Void volume fraction in the GIF2 simulation for a range of redshifts.}
  \label{voltable}
\end{table}

Table~\ref{voltable} shows the fraction of the simulation volume occupied by voids with radii larger than 
$r_{\rm min}$ in the GIF2 simulation. The volume fraction grows by approximately a factor of three between $z=3$ 
and 2, between $z=2$ and 1, and between $z=1$ and 0.  

\begin{figure}
\includegraphics[width=85mm]{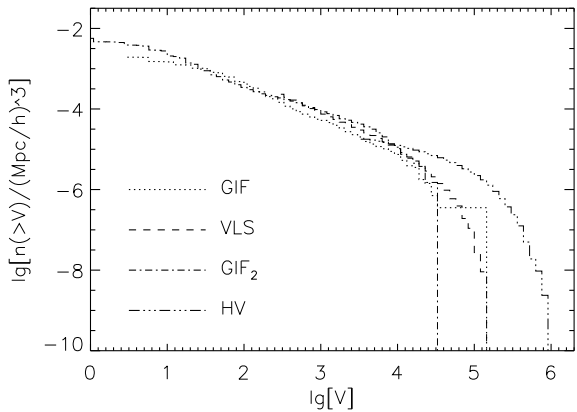}
\caption{Cumulative volume functions of voids at $z=0$ in the GIF (dotted), GIF2 (dot--dashed), VLS (dashed), 
         and HV simulations (dot--dot--dot--dashed).}
\label{VolumeFunctions}
\end{figure}

\begin{figure}
\includegraphics[width=85mm]{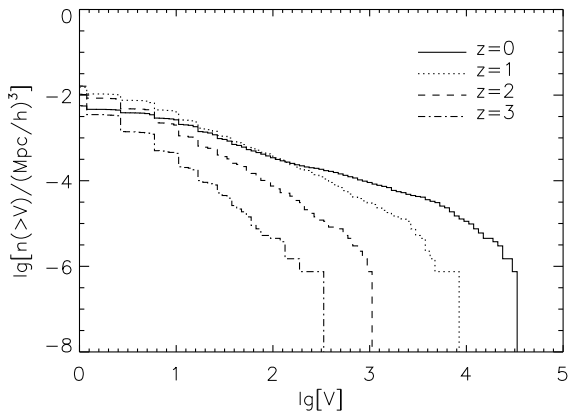}
\caption{Evolution of the the cumulative void volume fraction in the GIF2 simulation at $z=0$ (solid), $z=1$ 
         (dotted), $z=2$ (dashed), and $z=3$ (dot--dashed).}
\label{GIF2VolumeFunctions}
\end{figure}

\begin{figure}
\includegraphics[width=85mm]{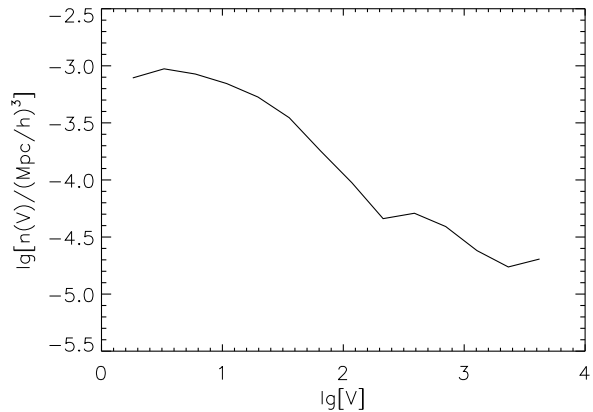}
\caption{Differential void volume function of voids in the GIF2 simulation at $z=0$.}
\label{GIF2VolumeFunctionDiff}
\end{figure}

\begin{figure}
\includegraphics[width=85mm]{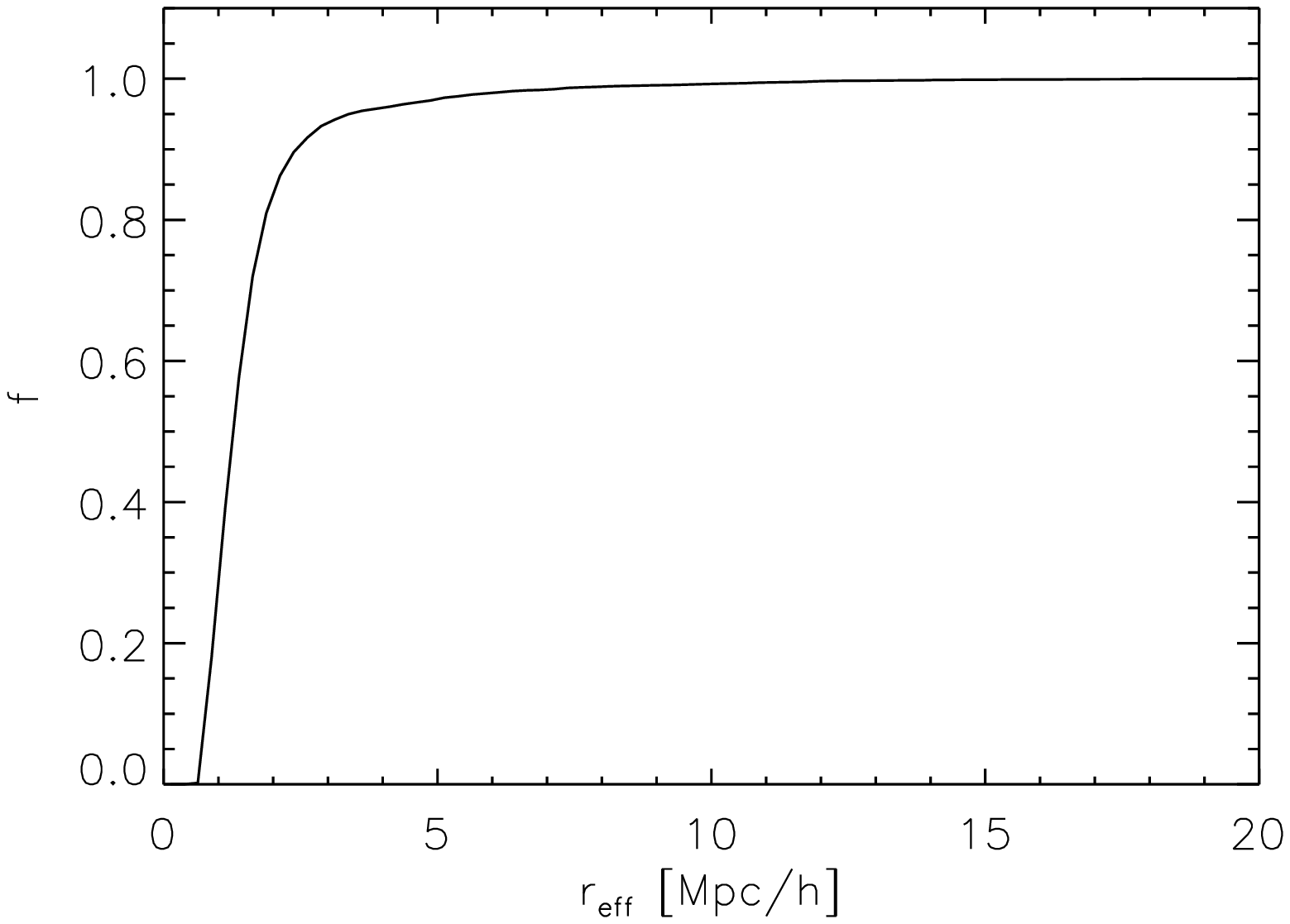}
\caption{Fraction of total void volume filled by voids of effective radius $r_{eff}$ or less in the GIF2 
         simulation at $z=0$. See text for discussion.}
\label{GIF2VolumeFraction}
\end{figure}

Figure \ref{VolumeFunctions} shows the number density of voids larger than a given volume $V$, as a function of 
$V$, at $z=0$ in the four simulations. The very big GIF void is clearly visible in this plot. It is almost as 
large as the largest voids in the much larger VLS simulation. The steps visible at small $V$ result from the 
discreteness of the grid. The agreement between the different simulations is really quite good. Of course, 
the Hubble Volume simulation contains by far the largest voids.

Figure~\ref{GIF2VolumeFunctions} shows how the cumulative volume function in the highest resolution simulation 
(GIF2) evolves. The evolution is smooth, and the volume functions of different redshifts cross each other. For 
example, there are more voids of volume 100~($h^{-1}$Mpc)$^3$ at $z=1$ than at $z=0$. This growth of voids is 
analogous to the hierarchical growth of haloes.  As time progresses, smaller haloes merge with one another to 
form larger haloes.  Here, smaller voids expand and merge with other voids to form larger ones. (We did not attempt 
to use our void catalogs to construct void merger trees.) This plot reflects what we have shown earlier in
the three sets of panels in Figure~\ref{GIFmosaic}. 

In Figure~\ref{GIF2VolumeFunctionDiff} we plot the differential void volume function of voids in the GIF2 
simulation at $z=0$. It is interesting to note that the distribution does not have a peak. 
When expressed as a function of $1.7/\sigma(m,z)$, the halo mass function is reasonably well fit by a universal 
form that is independent of redshift and power spectrum (Sheth \& Tormen 1999; Sheth, Mo \& Tormen 2001; Jenkins 
et al 2001). However, a similar rescaling of the void function (using $2.8/\sigma(m)$ -- see Sheth \& Van de 
Weygaert 2004), does not yield a universal curve. The failure to find a universal form contradicts excursion--set 
models of the sort that describe halos quite well, but is not in disagreement with models based on peaks in 
Gaussian random fields (Sheth \& Van de Weygaert 2004).

For the GIF2 simulation at $z=0$, in Figure~\ref{GIF2VolumeFraction} we plot the fraction of the total void volume 
filled by voids of effective radius $r_{eff}$ or less as a function of that radius. It is quite instructive to see 
that around half of the void volume is already filled at a radius of around 1.3\,$h^{-1}$\,Mpc, and voids with an 
effective radius of 2.5\,$h^{-1}$\,Mpc or less fill 90 percent of the void volume. In other words, even though the 
largest voids leave the strongest visual impression in images like Figure \ref{GIF2slice}, they only account for a 
small fraction of the total void volume.

It is not straightforward to compare these findings with results from investigations of voids in galaxy
catalogues. This is partly because of the difference in the void finding algorithms and mainly because
of the fact that galaxies are quite sparse tracers of the underlying density field. On a qualitative level,
our void size distribution agrees well with observed voids. For example, Hoyle \& Vogeley 2004 report void 
sizes comparable to our largest voids (the smallest voids they construct have radii of 10\,$h^{-1}$\,Mpc), 
with the numbers of voids steeply dropping with increasing radius.

\begin{figure*}
\includegraphics[width=170mm]{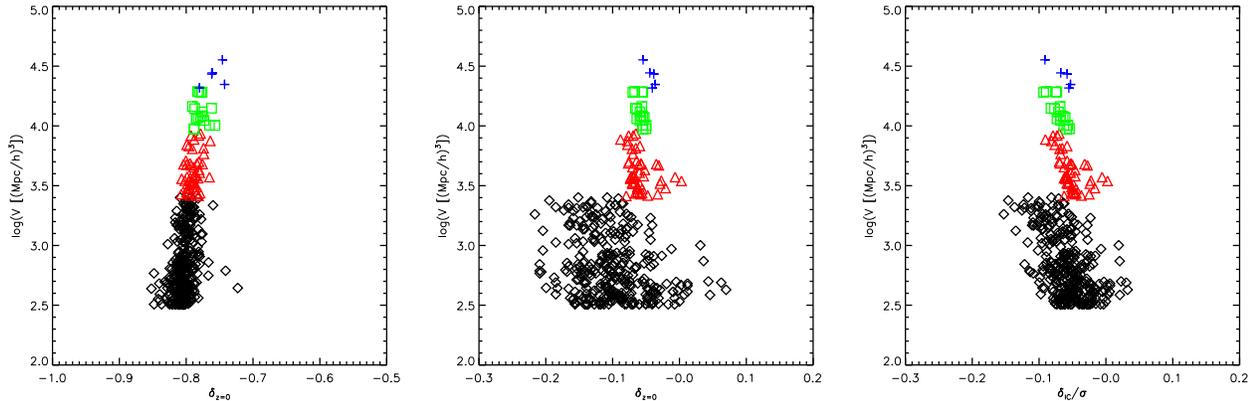}
\caption{Left--most panel: Void volumes versus void overdensities at $z=0$ in the GIF simulation. Center
         panel: Voids in the smoothed initial density field (GIF simulation). Plotted is the void volume at $z=0$ as a 
         function of the overdensity $\delta$ of the associated trough in the initial conditions. Different symbols 
         show different smoothing scales: 2.5\,$h^{-1}$\,Mpc (diamonds), 5.0\,$h^{-1}$\,Mpc (triangles), 
         7.5\,$h^{-1}$\,Mpc (squares), and 10.0\,$h^{-1}$\,Mpc (crosses). Right--most panel: Voids in the 
         smoothed initial density field (GIF simulation). Plotted is the void volume at $z=0$ as a 
         function of the re--scaled overdensity $\delta/\sigma$ of the associated trough in the initial conditions. 
         Different symbols show different smoothing scales: 2.5\,$h^{-1}$\,Mpc (diamonds), 5.0\,$h^{-1}$\,Mpc 
         (triangles), 7.5\,$h^{-1}$\,Mpc (squares), and 10.0\,$h^{-1}$\,Mpc (crosses).}
\label{VoidDelta}
\end{figure*}

\subsection{Voids In The Initial Density Field}\label{troughs}

Massive haloes in simulations are associated with higher peaks in the (smoothed) initial density field (Bardeen et al 
1986; Colberg et al 2000; Sheth \& Diaferio 2001). Voids are expected to form from initially underdense regions 
analogously to how clusters or haloes form from initially overdense regions. One might thus wonder if a similar 
correlation exists between voids and minima in the initial density field. We used the GIF simulations to study 
this correlation as follows.  

In the spherical evolution model, the mass associated with a void is a measure of the initial comoving radius of 
the region from which it formed: $R = (3m/4\pi\bar\rho)^{1/3}$.  Therefore, one might expect the void mass to 
correlate most strongly with the depth of the initial underdensity from which it formed, when the initial field 
is smoothed on a scale $R(m)$.  Since the voids in our sample enclose a large range of masses, we smoothed the 
initial ($z=49)$ density field using a set of Top Hat filters: 2.5, 5.0, 7.5, 10.0\,$h^{-1}$\,Mpc. We identified 
the minima in each smoothed field. That is, we identified those grid cells which were less dense than all twenty 
six of their neighbouring cells. We then compared the comoving positions of the minima identified on a smoothing 
scale with the locations of those voids whose $z=0$ sizes correspond to $R(m)$ -- recall how initially underdense
regions grow by a factor of 1.7 until present time. If there was more than one minimum inside a void we picked the 
deepest one. The density inside that cell was identified with the overdensity $\sigma$ of the trough. This method is 
analogous to how Colberg et al 2000 located peaks for clusters. What is more, voids evolve by expanding but not 
by moving. Thus, one expects to find the void centers in the initial conditions close to the void centers at present 
time. In this way, we associated voids with minima in the initial field. 

As it turns out, all voids larger than 4.25\,$h^{-1}$\,Mpc could be associated with a density minimum. It is 
interesting that associating a void with an initially underdense region does thus work much better than finding 
a peak for a cluster (see Colberg et al 2000). 

In the left--most panel of Figure~\ref{VoidDelta} we plot the void volumes at $z=0$ as a function of the 
void overdensities at $z=0$. The void overdensities scatter around the value of -0.8. Larger voids tend to be 
slightly less underdense. This is mainly due to the process of the merging of proto--voids. As will be seen in 
the following section, void density profiles rise very sharply towards the edges of the voids (see 
Figure~\ref{LargeDensityProfilesGIF}). Thus, when a smaller void is merged onto a larger one -- following the 
criteria outlines above -- one basically adds mainly parts of the outer region of the smaller void. Once the 
overdendity of the resulting void is computed this void will have a slightly higher overdensity than the two 
original voids.

The center panel of Figure~\ref{VoidDelta} shows the $z=0$ void volume as a function of the overdensities of the 
associated troughs in the initial conditions. As discussed above, we used a set of smoothing scales and grouped 
the voids into categories covered by the corresponding scale. In principle, for each void one would want to apply 
a smoothing scale that corresponds exactly to the void volume. Since we did not do that we end up with clearly 
visible steps in the plot.

If one re--scales overdensities of the associated troughs (compare Sheth \& Diaferio 2001 for the analogous 
procedure for haloes) the plot gets tighter. The right--most panel of Figure~\ref{VoidDelta} shows the $z=0$ 
void volume as a function of $\delta/\sigma$ of the associated troughs. The different sets are still visible 
but now they lie on top of each other.

\subsection{Void Density Profiles}\label{profiles}

Navarro et al (1996) have argued that CDM haloes have a universal density profile. In this section, we argue that the 
same holds true for voids, in qualitatively agreement with Van Kampen \& Van de Weygaert (1993).  

\begin{figure}
\includegraphics[width=85mm]{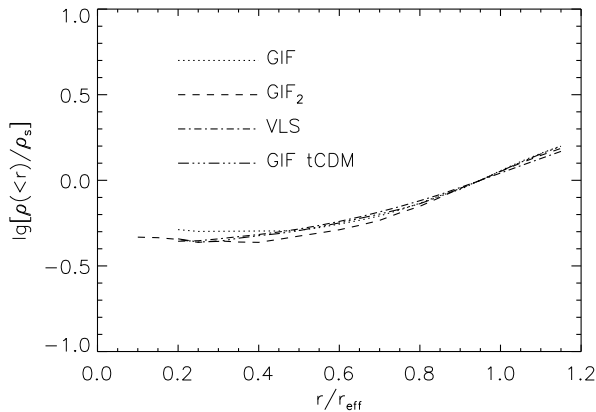}
\caption{Enclosed density in $z=0$ voids as a function of radius for the GIF (dotted line), GIF2 (dashed line), and 
         VLS (dot--dashed line) simulations. For each simulation, the rescaled profiles of voids with radii larger 
         than $5\,h^{-1}$\,Mpc were averaged (i.e., radii and densities were scaled by the effective radius and the 
         enclosed density at the effective radius, before averaging). In addition, the results from the $\Omega=1$ 
         $\tau$CDM GIF simulation are shown (three dots--dashed line). Curves are truncated at small radii because of 
         numerical resolution limits.}
\label{CumDensityProfiles}
\end{figure}

Using our samples from the GIF, GIF2, and VLS simulations, we have computed the mass profiles of voids, using the 
actual particles in the voids instead of the smoothed density grid (we found the difference was important).  
Because of the different lower thresholds of the samples, we only compute void profiles for voids that have effective 
radii of $5\,h^{-1}$\,Mpc or more. Figure~\ref{CumDensityProfiles} shows the averaged enclosed density in $z=0$ voids 
as a function of radius. For each void, we re--scaled the length scales by dividing by the effective radius, and we 
re--scaled densities by dividing by the enclosed density at the effective radius. We truncated the profiles at small 
radii, where numerical resolution effects begin to dominate (these will be discussed in more detail below). For almost 
the entire range, the average density profiles of voids in the three simulation sets agree very well. We also computed
density profiles for the $\Omega=1$ $\tau$CDM GIF simulation. These agree with the profiles of the $\Lambda$CDM 
simulations. This finding indicates that the form of void density profiles is indeed universal.

The different mass resolution of the three simulations affects the profiles in a systematic way: the higher the 
resolution, the lower the profile. This effect is strongest at small radii. The mass resolution also affects the 
centers of the voids. For example, in the VLS simulation, a single particle in a sphere of radius $1\,h^{-1}$\,Mpc 
corresponds to an overdensity of $-0.8$. Therefore, we cannot resolve the density profiles in the innermost regions. 
If we plot the profiles all the way to the centers we find that the profiles all rise -- individual particles 
contribute too much mass. Therefore, we truncate the profiles in the void centers. We cross--checked the effect 
of mass resolution by down--sampling the GIF2 simulation and producing void profiles. The down--sampled simulation 
shows the trend visible in Figure~\ref{CumDensityProfiles}. Figure~\ref{CumDensityProfiles} is very encouraging: 
except for the effect of mass resolution, there are no systematic differences in the void samples. 

\begin{figure}
\includegraphics[width=85mm]{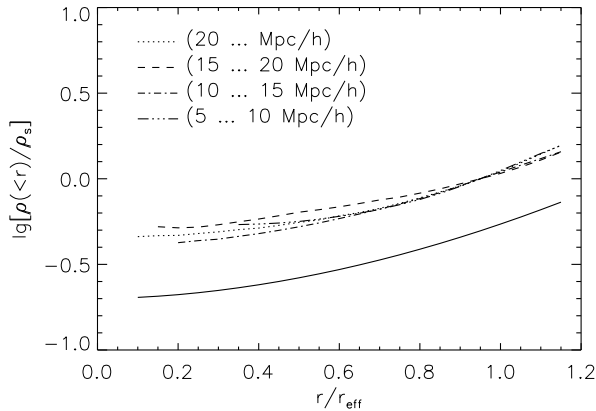}
\caption{Enclosed density profiles in the VLS simulation. The voids were divided into four samples with void radii 
         between 5 and $10h^{-1}$Mpc (dashed-three-dots), 10 and $15h^{-1}$Mpc (dot dashed), 15 and $20h^{-1}$Mpc 
         (dashed), and voids with radii larger than $20h^{-1}$Mpc (dotted). The radius and enclosed mass of each 
         void was rescaled by the effective radius and the effective mass, and these rescaled profiles were averaged 
         over all voids. We have truncated the curves at small radii because of numerical resolution limits of the 
         simulations. Solid line shows equation~(\ref{fitcumdens}), shifted downwards by a factor of two.}
\label{CumDensityProfilesVLS}
\end{figure}

\begin{figure}
\includegraphics[width=85mm]{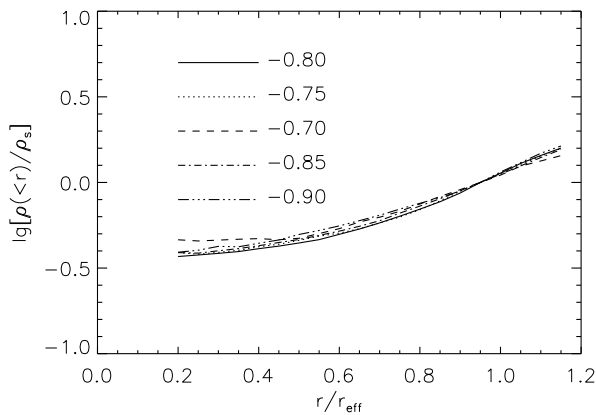}
\caption{Cumulative density profiles for different void overdensity definition thresholds in the GIF simulation.}
\label{SystematicDensityProfiles}
\end{figure}

Figures~\ref{CumDensityProfilesVLS}, shows the scaled enclosed density profiles of voids in the VLS simulation.  
The curves lie fairly nicely on top of each other. The cumulative profile shown in Figure~\ref{CumDensityProfilesVLS} 
is quite well described by 
\begin{equation}
\rho(<r)/\rho(r_{\rm eff}) = \frac{\exp[(r/r_{\rm eff})^{1.85}]}{2.5}\,.
\label{fitcumdens}
\end{equation}
The fits only start to deviate somewhat beyond $r/r_{eff} = 1$. (Although the density run around halos is usually 
studied using the differential profile, void centers have fewer particles, so we have chosen to fit to the cumulative 
profile instead.) Our choice of an exponential profile is motivated by Van de Weygaert \& Van Kampen 1993 who noted that
an exponential profile provided a very good fit to their voids.

Figure \ref{SystematicDensityProfiles} shows the density profiles of voids in the GIF simulation with a range of 
values for the mean void overdensity. As can be seen, varying the overdensity threshold in the range chosen here 
does not systematically alter the density profiles.
 
\begin{figure*}
\includegraphics[width=170mm]{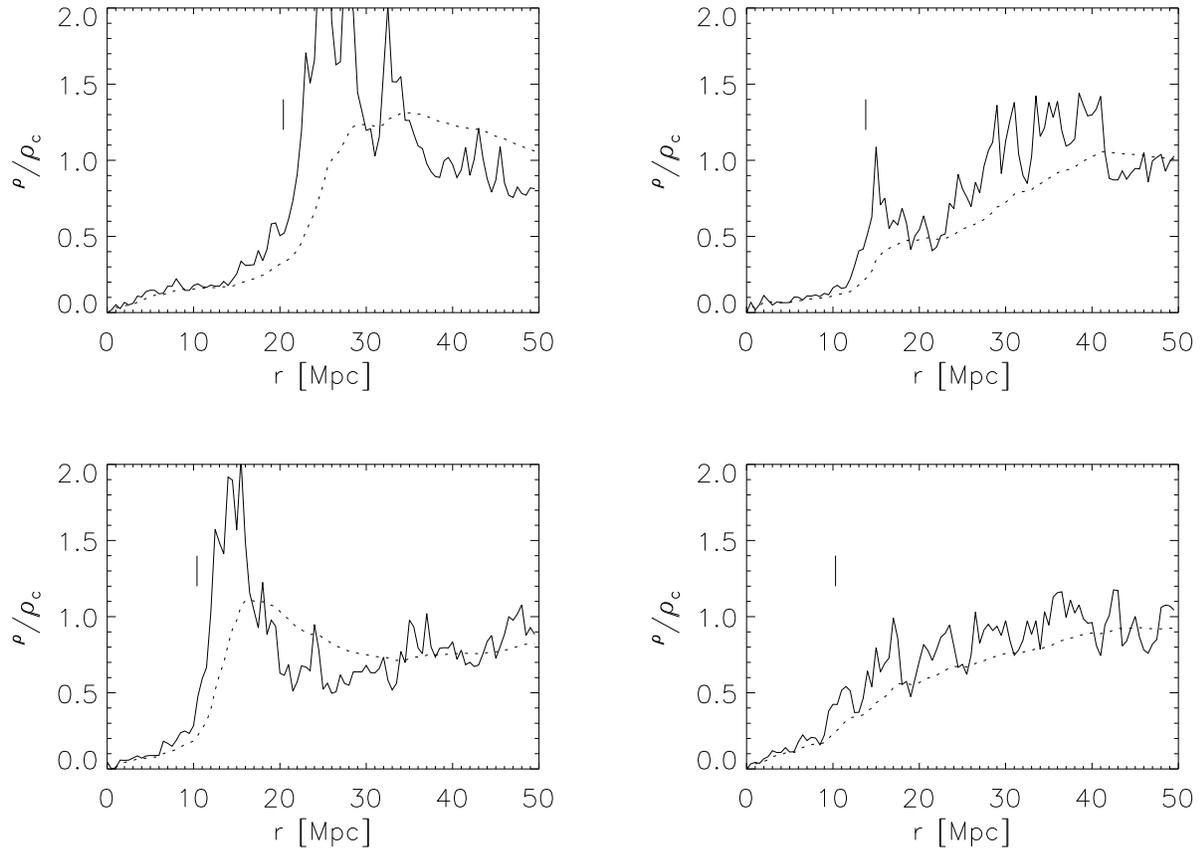}
\caption{Density profiles of four voids from the GIF simulation. Solid and dashed lines show the density and enclosed 
         density as a function of distance from the void center.  Vertical line shows the effective radius of each void.}
\label{LargeDensityProfilesGIF}
\end{figure*}

Figure \ref{LargeDensityProfilesGIF} shows density profiles of four GIF voids going out to a distance of $50h^{-1}$Mpc 
from their centers. The void edges are marked with a small vertical line. Although there are some variations in the 
profiles, all voids have very sharp edges. The densities peak at the effective radius, and the enclosed densities 
rise above the threshold. This is consistent with the visual impressions of voids discussed earlier, where one sees 
that voids are very well defined by the haloes which populate their boundaries. It also agrees qualitatively with 
the results in Van de Weygaert \& Van Kampen (1993). What is more, Benson et al (2003) and Hoyle \& Vogeley 2004 see 
similar behaviour for voids outlined by the galaxy distribution in a semi-analytic galaxy formation model and in the 2dFGRS,
respectively (see Figure 11 in Benson et al (2003) and Figure 4 in Hoyle \& Vogeley 2004).

\subsection{The Void Halo Mass Function} \label{massfunction}

Gottl\"ober et al (2003) investigated the $z=0$ void halo mass function using a set of high--resolution simulations of 
individual voids. They find that both the normalization and the shape of the cumulative mass function are different 
from those of the non--void halo mass function.  Their measurements are in qualitative agreement with models for 
this dependence by Mo \& White (1996) and Sheth \& Tormen (2002), although there are differences in detail. Also
see Patiri et al (2004) who used the simulations run by Gottl\"ober et al (2003) to model mass functions in voids. 

For our study of the mass function, we use the GIF2 simulation, which has the highest mass resolution. We identify 
haloes using a friends--of--friends (fof) group finder with a linking length of 0.2 times the mean interparticle 
separation, and require that haloes have at least 10 particles. At $z=0$, we find void haloes by picking those 
haloes whose centres--of--mass lie within a void.\footnote{For this part of this work we do not use our estimates 
of the void centers and effective radii. Instead, void haloes are defined as those which lie within a void--volume, 
however complex its shape.}  We then mark those particles that are in a void at $z=0$ and run the fof group finder 
on them at earlier redshifts. This means that we do not require that $z=0$ void haloes be located inside a void at 
earlier times. Our choice is dictated by the fact that the void volume fraction evolves rapidly (c.f. 
Section~\ref{VoidVolumeFunction}); choosing only haloes that are inside voids at early times would reduce the size 
of our high redshift halo samples significantly.  Thus, what we are really showing is the mass function of the 
high-$z$ projenitors of halos which are in voids at $z=0$.  

\begin{figure}
\includegraphics[width=85mm]{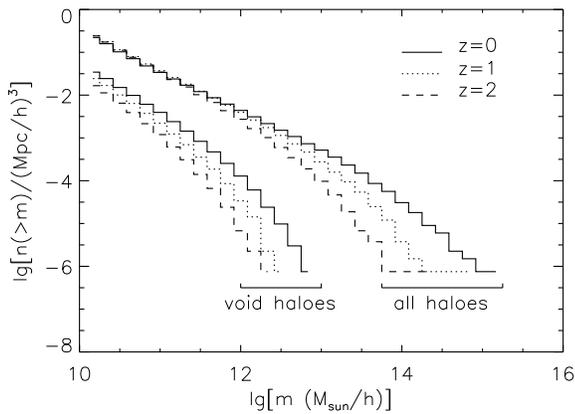}
\caption{Mass function of haloes in the GIF2 simulation. Different curves show the mass function of all haloes, 
         whatever their surrounding environment, and the mass function of those haloes whose particles lie in a void 
         at $z=0$.}
\label{MassFunctionGIF2_2}
\end{figure}

\begin{figure}
\includegraphics[width=85mm]{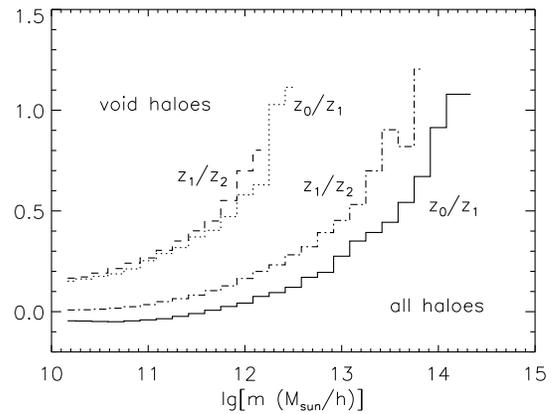}
\caption{Ratios of the mass functions of haloes in the GIF2 simulation in different environments. The solid 
         ($z=0/z=1$) and dot--dashed ($z=1/z=2$) and the dotted ($z=0/z=1$) and dashed lines ($z=1/z=2$) are 
         for all haloes and to void haloes, respectively.}
\label{MassFunctionGIF2_3}
\end{figure}

Figure~\ref{MassFunctionGIF2_2} compares the mass function of all haloes with that of haloes whose particles lie in 
a void at $z=0$.The plot indicates that haloes that end up in a void at $z=0$ -- probably located at the very edges 
of a void -- {\it at any fixed mass\/} undergo slightly more evolution than haloes with the same mass elsewhere. 
Figure~\ref{MassFunctionGIF2_3} shows this point a little bit more clearly by plotting the ratios of the 
mass fuctions shown in Figure~\ref{MassFunctionGIF2_2} for $z=0/z=1$ and $z=1/z=2$. Note that if you look at 
all haloes, for small halo masses there are less haloes at later redshift ($z=0$) than at the earlier redshift 
($z=1$).

The small simulation volume and the resulting modest halo sample sizes do not allow more detailed studies of this. 
We will re--address the void halo mass function in a later study that will make use of a much larger simulation.  

\subsection{The Spatial Distribution of Voids} \label{corrfct}

\begin{figure}
\includegraphics[width=85mm]{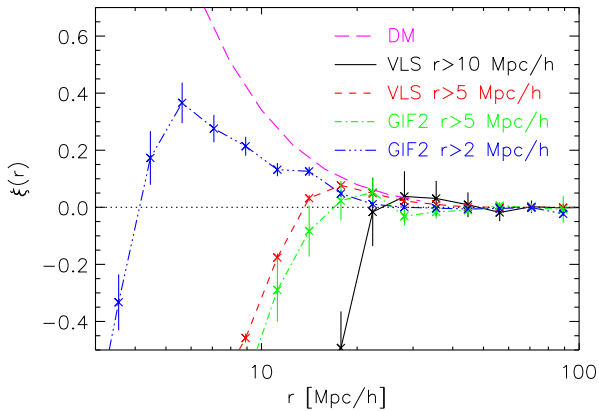}
\caption{The two--point correlation function $\xi(r)$ for voids of radius $r>2$\,Mpc/$h$ (from GIF2, dash three
         dots), $r>5$\,Mpc/$h$ (dot dash from GIF2 and dashed from VLS), and $r>10$\,Mpc/$h$i (solid line).
         The long dashed line shows the dark matter correlation function.}
\label{CorrFunction}
\end{figure}

In this Section, we want to investigate the spatial clustering of voids in detail by computing the two--point 
correlation function of the voids. The correlation function of massive haloes depends strongly on halo mass 
(Mo \& White 1996; Sheth \& Tormen 1999), so it is interesting to see if voids show analogous trends.  

To study a large range of void sizes, we measured the correlation function of voids centres in the VLS and GIF2 
simulations. Figure \ref{CorrFunction} shows the results for voids with radii $R>2$\,Mpc/$h$ (three dots dash)
and $R>5$\,Mpc/$h$ (dot dash)
from GIF2, and with $R>5$\,Mpc/$h$ (short dash) and $R>10$\,Mpc/$h$ (solid) from VLS. At separations smaller 
than $\sim 2R$, the 
void correlation functions tend to $-1$.  This is a consequence of volume exclusion: for the purposes of this 
statistic, voids are like hard spheres -- they do not overlap. At larger radii, there is some evidence that the 
larger voids are slightly more clustered, consistent with a linear peaks-bias based model. (E.g., note the 
similarity between Figure \ref{CorrFunction} and Figures~2 and~3 in Sheth \& Lemson 1999.)  However, the scales 
on which volume exclusion effects are no longer important, and on which the linear bias model may apply are 
sufficiently large that the amplitude of the unbiased (dark matter) correlation function (long dash) is small.  Hence, the 
actual amplitude of the void correlation function is never large compared to unity.  This provides considerable 
justification for the accuracy of Van de Weygaert's (2002) Poisson Voronoi based models of the matter distribution.

\section{Conclusions} \label{conclusions}

We studied the properties of voids in a set of large high-resolution N-body simulations of the $\Lambda$CDM 
cosmology. We defined voids as spherical or elliptical regions of space with a mean overdensity of $-0.8$. 
With this definition we found almost 80,000 voids with radii larger than $10h^{-1}$Mpc in the Hubble Volume 
simulation. Those voids fill the volume approximately uniformly.

The void volume functions of the different simulations agree well. The largest void in the HV simulation has a 
radius of $\sim 55.9h^{-1}$Mpc. It is quite interesting that this is fairly close to the size of the famous 
void in the region of Bo\"otes found by Kirschner et al (1981).

The GIF simulation appears to harbor an abnormally large void, given the small size of the simulation box. 
There are more smaller voids at earlier times than at later times (Fig.~\ref{GIF2VolumeFunctions}). Claims 
that CDM cosmologies do not form large enough voids can thus be put to rest. In addition, as our voids are 
defined through their mean overdensity we also show that CDM voids do not contain too much matter.

Voids very clearly correspond to troughs in the smoothed initial density field (right--most panel of 
Fig~\ref{VoidDelta}). This point is particularly interesting in the light of Colberg et al (2000)'s result 
for the correspondence between clusters and peaks: They found that not all clusters could be associated 
with peaks. For voids, the idea that the initial density field contains the seeds of $z=0$ objects can be 
verified much more successfully.

When appropriately rescaled, voids appear to have a universal density profile (eq.~\ref{fitcumdens}). The 
void density profiles rise steeply at the edges of voids.  Voids are thus very well defined in terms of their 
densities.

In agreement with the results reported by Gottl\"ober at al (2003) we find that the mass function of haloes 
in voids is different from that in regions of average density 
(Fig.~\ref{MassFunctionGIF2_2}). We also find that the mass function of haloes which end up in $z=0$ voids 
evolved somewhat more rapidly than the mass function of all haloes. However, even the simulation with the 
highest mass resolution in our set just barely reaches down to the mass range of void haloes. For a detailed 
investigation of formation times one would need simulations with an even higher mass resolution, such as those 
used by Gottl\"ober et al (2003).  This is the subject of work in progress.

While there is some evidence that larger voids are slightly more clustered on scales larger than
$\sim 20\,h^{-1}$\,Mpc the actual amplitude of the void correlation function is very small. This finding
supports the use of Poisson Voronoi based models of the matter distribution (Van de Weygaert 2002). 

\section*{Acknowledgments}

The simulations discussed here were carried out as part of the Virgo Consortium programme, on the Cray T3D/Es at 
the Rechenzentrum of the Max--Planck-Gesellschaft in Garching, Germany, and at the Edinburgh Parallel Computing Center.
We are indebted to the Virgo Supercomputing Consortium for allowing us to use these simulations. We thank Andy 
Connolly, Rupert Croft, Darren Croton, Alexander Knebe, Simon Krughoff, Volker Springel, Felix St\"ohr, and
especially Rien van de Weygaert for discussions, encouragement, and suggestions.
We also thank Adrian Jenkins for providing some of the simulations 
and for comments on an earlier draft of the work. JMC acknowledges partial support through research grant ITR AST0312498. 
NY acknowledges support from JSPS Special Research Fellowship SPD-0302674.

\label{lastpage}


\begin{thebibliography}{99}
\bibitem{a1} Aikio J., M\"ah\"onen P., 1998, ApJ, 497, 534
\bibitem{b1} Benson A.J., Hoyle F., Torres F., Vogeley M.S., 2003, MNRAS, 340, 160
\bibitem{b3} Bertschinger E., 1985, ApJ, 58, 1
\bibitem{b2} Blumenthal G.\ R., Da Costa L,\ N., Goldwirth D.\ S., Lecar M., Piran T., 1992, ApJ, 388, 234
\bibitem{b4} Bond J.\ R., Efstathiou G., 1984, ApJ, 285, 45
\bibitem{c1} Colberg J.\ M., White S.\ D.\ M.\, MacFarland T.\ J., Jenkins A., Pearce F.\ R., Frenk C.\ S., 
             Thomas P.\ A., Couchman H.\ M.\ P. (The Virgo Consortium), 2000, MNRAS, 313
\bibitem{c2} Croton D.\ J., et al., astro-ph/0407537
\bibitem{d1} da Costa L.N., Geller M.J., Pellegrini P.S., Latham D.W., Fairall A.P., Marzke R.O.,
             Willmer C.N.A., Huchra J.P., Calderon J.H., Ramella M., Kurtz M.J., 1994, ApJ, 424, 1
\bibitem{d2} Davis M., Efstathiou G., Frenk C.S., White S.D.M., 1985, ApJ, 292, 371
\bibitem{d3} Dubinski J., Da Costa L,\ N., Goldwirth D.\ S., Lecar M., Piran T., 1993, ApJ, 410, 458
\bibitem{e1} Einasto M., Tago E., Jaaniste J., Einasto J., Andernach H., 1997, A\&AS, 123, 119
\bibitem{e2} Einasto J., Saar E., Einasto M., Freudling W., Gramann M., 1994, ApJ, 429, 465
\bibitem{e3} El-Ad H., Piran T., 1997, ApJ, 491, 421
\bibitem{e4} El--Ad H., Piran T., da Costa L.N., 1997, MNRAS, 287, 790
\bibitem{e5} Evrard A.\ E., MacFarland T.\ J., Couchman H.\ M.\ P., Colberg J.\ M., Yoshida N., White S.\ D.\ M., 
             Jenkins A., Frenk C.\ S., Pearce F.\ R., Peacock J.\ A., Thomas P.\ A., 2002, ApJ, 573, 7
\bibitem{f1} Friedmann Y., Piran T., ApJ, 548, 1
\bibitem{g1} Gao L., White S.\ D.\ M., Jenkins A., Stoehr F., Springel V., astro-ph/0404589
\bibitem{g2} Gardner J.P, 2001, ApJ, 557, 616
\bibitem{g3} Geller M.J., Huchra J.P., 1989, Science, 246, 857
\bibitem{g4} Goldberg D.M., Vogeley M.S., astro-ph/0307191
\bibitem{g5} Gottl\"ober S., Lokas E., Klypin A., Hoffman Y., 2003, MNRAS, 344, 715
\bibitem{h1} Hoyle F., Vogeley M.\ S., 2002, ApJ, 566, 641
\bibitem{h2} Hoyle F., Vogeley M.\ S., 2004, ApJ, 607, 751
\bibitem{i1} Icke V., 1984, MNRAS, 206, 1
\bibitem{j1} Jenkins A., Frenk C.\ S., Pearce F.\ R., Thomas P.\ A., Colberg J.\ M., White S.\ D.\ M., 
             Couchman H.\ M.\ P., Peacock J.\ A., Efstathiou G., Nelson, A.\ H. (The Virgo Consortium), 
             1998, ApJ, 499, 20
\bibitem{j2} Jenkins A., Frenk C.\ S., White S.\ D.\ M., Colberg J.\ M., Cole S., Evrard A.\ E., 
             Couchman H.\ M.\ P., Yoshida N., 2001, MNRAS, 321, 372 
\bibitem{k1} Kauffmann G., Fairall A.\ P., 1991, MNRAS, 248, 313
\bibitem{k2} Kauffmann G., Melott A.\ L., 1992, ApJ, 393, 415
\bibitem{k3} Kauffmann G., Colberg J.\ M., Diaferio A., White S.\ D.\ M., 1999, MNRAS, 303, 188
\bibitem{k4} Kirshner R.\ P., Oemler A., Schechter P.\ L., Shectman, S.\ A., 1981, ApJ, 248, 57
\bibitem{l1} Lacey C. Cole, S., 1994, MNRAS, 271, 676
\bibitem{l3} Lindner U., Einasto M., Einasto J., Freudling W., Fricke K., Lipovetsky V., Pustilnik S.,
             Izotov Y., Richter G., 1996, A\&A, 314, 1
\bibitem{l4} Lindner U., Einasto J., Einasto M., Freudling W., Fricke K., Tago E., 1995, A\&A, 301, 329
\bibitem{l5} Little B., Weinberg D.\ H., 1994, MNRAS, 267, 605
\bibitem{m1} Mathis H., White S.D.M., 2003, MNRAS, 337, 1193
\bibitem{m2} Menard B., Hamana T., Bartelmann M., Yoshida N., 2003, A\&A, 403, 817
\bibitem{m3} M\"uller V., Arbabi--Bidgoli S., Einasto J., Tucker D., 2000, MNRAS, 318, 280
\bibitem{n1} Navarro J. F., Frenk C. S., White S. D. M., 1996, ApJ, 462, 563
\bibitem{p1} Patiri S. G., Betancort--Rijo J. E., Prada F., astrop--ph/0407513
\bibitem{p2} Peebles P.\ J.\ E., 2001, ApJ, 557, 495
\bibitem{p3} Plionis M., Basilakos S., 2002, MNRAS, 330, 399
\bibitem{r1} Reed D., Gardner J., Quinn T., Stadel J., Fardal M., Lake G., Governato F., 
             astro-ph/0301270
\bibitem{r2} Rojas R.\ R., Vogeley M.\ S., Hoyle F., Brinkmann J., astro-ph/0307274
\bibitem{s2} Schmidt J.\ D., Ryden B.\ S., Melott A.\ L., 2001, ApJ, 546, 609
\bibitem{s6} Seljak, U., Zaldarriaga M., 1996, ApJ, 469, 437
\bibitem{s1} Shectman S.\ A., Landy S.\ D., Oemler A., Tucker D.\ L., Lin H., Kirshner R.\ P., Schechter P.\ L.,
             1996, ApJ, 470, 172
\bibitem{s3} Sheth, R.\ K., Lemson G., 1999, MNRAS, 304, 767
\bibitem{s4} Sheth R.\ K., Diaferio A., 2001, MNRAS, 322, 901
\bibitem{s5} Sheth R.\ K., Van de Weygaert R., 2004, MNRAS, 350, 517
\bibitem{v1} Van de Weygaert R., Van Kampen E., 1993, MNRAS, 263, 481
\bibitem{y1} Yoshida N., Sheth R.\ K., Diaferio A., 2001, MNRAS, 328, 669
\end{thebibliography}
\end{document}